\documentclass[12pt]{article}
\usepackage{graphicx}

\usepackage{scicite}

\usepackage{times}

\newenvironment{sciabstract}{%
\begin{quote} \bf}
{\end{quote}}

\newcounter{lastnote}
\newenvironment{scilastnote}{%
\setcounter{lastnote}{\value{enumiv}}%
\addtocounter{lastnote}{+1}%
\begin{list}%
{\arabic{lastnote}.}
{\setlength{\leftmargin}{.22in}}
{\setlength{\labelsep}{.5em}}}
{\end{list}}

\title{Two white dwarfs with oxygen-rich atmospheres}

\author{B.T. G\"ansicke$^1$, D. Koester$^2$,
J. Girven$^1$, T.R. Marsh$^1$, D. Steeghs$^1$\\
\normalsize
$^{1}$ Department of Physics, University of Warwick, Coventry CV4 7AL, UK\\
\normalsize
$^{2}$ Institut f\"ur Theoretische Physik und Astrophysik, University of Kiel,
24098 Kiel, Germany}

\date{}

\begin{document} 


\baselineskip24pt


\maketitle



\begin{sciabstract}
Stars with masses in the range $7-10\,M_\odot$ end their lives either
as massive white dwarfs or weak type II supernovae, and there are only
limited observational constraints of either channel. Here we report
the detection of two white dwarfs with large photospheric oxygen
abundances, implying that they are bare oxygen-neon cores and that
they may have descended from the most massive progenitors that avoid
core-collapse.
\end{sciabstract}



White dwarfs represent the endpoints of stellar evolution for the
overwhelming majority of all stars. Most white dwarfs in the Galaxy
have carbon-oxygen (CO) core compositions, being descendants of low to
intermediate mass stars that underwent hydrogen and helium core
burning. Stars with initial masses $7M_\odot\le M \le 10M_\odot$ will
reach sufficiently high core temperatures to proceed to carbon
burning, and produce either oxygen-neon (ONe) core white dwarfs, or
undergo a core-collapse supernova (SN\,II) via electron capture on the
products of carbon burning \cite{nomoto84-1}. The exact outcome of
stellar evolution in this mass range depends critically on the
detailed understanding of mass loss \cite{hegeretal03-1}, the relevant
nuclear reaction rates and the efficiency of convective mixing in the
stellar cores \cite{stranieroetal03-1}. Some observational constraints
on stellar models come from analysis of SN\,II progenitors
\cite{smarttetal09-1}, which suggest a lower limit on the progenitor
masses of $8^{+1.0}_{-1.5}M_\odot$. A linear extrapolation of the
observed relation between the masses of white dwarfs and their
progenitors \cite{williamsetal09-1} up to the maximum mass of white dwarfs
leads to a broadly consistent result.

Information about the core compositions of white dwarfs, in particular
of those descending from the highest possible masses, would help to
test and improve the theory of stellar evolution. Unfortunately,
almost all white dwarfs have hydrogen and/or helium envelopes that,
while low in mass, are sufficiently thick to shield the core from
direct view. Asteroseismology has the potential to unveil their core
composition \cite{corsicoetal04-1}, but observational studies
attempting to exploit this potential remain ambiguous [see e.g. the
  case of GD\,358, \cite{metcalfeetal01-1, fontaine+brassard02-1,
    stranieroetal03-1}].  Core material can be directly detected in
the photospheres of a small number of stars \cite{werner+herwig06-1}
that underwent a very late thermal pulse during their asymptotic giant
branch (AGB) evolution, ejecting a large fraction of the envelope
\cite{herwigetal99-1} and leaving a white dwarf with
only a thin layer of helium. Examples of hydrogen-deficient post-AGB
stars are the hot PG1159 objects, white dwarfs with helium-rich (DB)
atmospheres and cool carbon-rich (DQ) white dwarfs. The most extreme
case known to date is 1H1504+65, which is hydrogen and helium
deficient, representing a hot, naked CO surface \cite{werneretal04-1}.

Recently, the Sloan Digital Sky Survey (SDSS), through its
comprehensive spectroscopic snapshot of the Galactic stellar
population, has revealed a small class of hydrogen-deficient white
dwarfs \cite{liebertetal03-2}. Their spectra are consistent with
nearly pure carbon atmospheres \cite{dufouretal07-1}. It has been
suggested that these ``hot'' DQ white dwarfs represent the
evolutionary link between objects such as PG1159 and 1H1504+65, the DB
white dwarfs, and the cool DQ white dwarfs \cite{dufouretal08-1,
  althausetal09-1}.

Until now, all known white dwarfs for which photospheric oxygen and
carbon abundances have been determined have abundance ratios
$\mathrm{O/C}\le1$. Some stellar models predict that the most massive
stars avoiding core-collapse will result in ONe white dwarfs with very
low carbon abundances \cite{ibenetal97-1}. Should such a core lose its
hydrogen envelope, an extremely oxygen-rich spectrum would be
expected. Here we present the results of a search for white dwarfs
with large photospheric oxygen abundances within the SDSS
spectroscopic Data Release (DR)~7 \cite{abazajian09-1}.

We selected all spectroscopic objects in DR7 that fall within the
$ugriz$ color space of the white dwarfs from
\cite{eisensteinetal06-1}.  We then subjected these spectra to an
automatic measurement of the equivalent widths [see
  \cite{gaensickeetal07-1} for details of the method] of the O\,I
616\,nm and 778\,nm multiplets, which are the strongest O\,I lines
expected in a cool, oxygen-rich atmosphere. Another strong O\,I
multiplet is located at 845\,nm, however, this region can be affected
by residuals from the night sky line subtraction, and we therefore did
not include it in our procedure.  As an additional constraint, we
required the spectra to have a signal-to-noise ratio $\ge10$ in the
regions around the two O\,I lines, resulting in the analysis of 25639
SDSS spectra \cite{note1}.

We found 1000 spectra with a formal $>4\sigma$ detection of both O\,I
lines, and visual inspection of these spectra showed that 998 were
erroneously flagged by our automated routine because of poor data
quality. Only two objects with genuine O\,I 616\,nm and 778\,nm
absorption, SDSS\,J092208.19+292810.9 and SDSS\,J110239.69+205439.4,
survived our scrutiny. Both have high proper motions, 0.275"/yr and
0.163"/yr, respectively \cite{lepine+shara05-1}, consistent with them
being nearby low-luminosity blue stars. Aside from the O\,I 616\,nm
and 778\,nm absorption lines, the spectrum of SDSS\,0922+2928
(Fig.\,1) is typical of a cool DQ white dwarf, with strong C$_2$ Swan
bands in the blue, superimposed by a number of atomic carbon lines.
In contrast SDSS\,1102+2054 is a unique white dwarf \cite{note2} with
a photospheric spectrum totally dominated by absorption lines of O\,I
(Fig.\,1). The only other noticeable features are some weak lines of
atomic C\,I as well as H$\alpha$ and H$\beta$.

We determined the effective temperatures of the two white dwarfs by
fitting their SDSS $ugriz$ photometry with two grids of theoretical
atmosphere models: one for a pure He composition [Supporting Online
  Material (SOM), Sect.\,1], and one including H, C, and O as well
(SOM, Sect.\,2). The white dwarf temperatures and chemical abundances
determined from these fits depend only mildly on the assumed surface
gravity $\log g$ and, because their masses are not constrained by the
available observations, we fixed $M=0.6M_\odot$, corresponding to the
canonical value of $\log g=8$ [\cite{dufouretal05-1}, SOM Sect.\,2].

Independent of the details of the atmospheric composition, the
photometry suggests that both white dwarfs have temperatures in the
range 8000--9000\,K (Table\,2). The implied distances depend strongly
on the assumed surface gravity via the radius, here we used the
mass-radius relation of \cite{wood95-1}. We used the effective
temperatures (and assumed surface gravity) to calculate synthetic
spectra for a final determination of atmospheric abundances from the
absorption lines in the SDSS spectroscopy (SOM, Sect.\,2).

For SDSS\,1102+2054, we adopted $T_\mathrm{eff}=8150$\,K and $\log
g=8$ and iterated the abundances of H, C, and O, until we obtained a
reasonable fit to the line strengths. Because this was not possible
for all oxygen lines we concentrated on the strong lines at 778 and
845\,nm, which originate from the lowest lying levels of all optical
lines. These lines depend only weakly on temperature and are best
suited for a determination of abundances.  The best compromise was
achieved for the abundances in Table\,3.

However, for this model the carbon multiplet near 712\,nm and the
834\,nm line, H$\alpha$, and many O\,I lines from higher levels are
much too broad, indicating too high a pressure in the atmosphere. In
addition, O\,I lines from the higher levels are much too weak.  The
most likely solution of this discrepancy is to assume a higher
temperature, which increases the strength of the high-excitation lines
compared to the strong multiplets (SOM, Sect.\,2). A reasonable
compromise was achieved with $T_\mathrm{eff}=10500$\,K, $\log g = 8$
(Table\,3). The remaining differences in the absorption lines between
the model and observation are likely within the uncertainties of the
atomic data and in particular the van der Waals broadening constants
(SOM, Sect.\,2).

This model, however, is in significant disagreement with the
temperature derived from the SDSS photometry.  One plausible cause for
the discrepancy between the temperatures determined from our fits to
the photometry and to the spectroscopy are uncertainties in the
ultraviolet opacities of an atmosphere of such peculiar composition,
which strongly affect the slope of the optical continuum
[\cite{dufouretal08-1}, SOM Sect.\,1].

The atmospheric parameters of SDSS\,1102+2054 thus remain somewhat
uncertain. Nevertheless, when varying the parameters within the range
demanded by photometry, line strengths and widths, the C and O
abundances did not change more than factors of $\sim3$. Thus the O/C
abundance ratio in SDSS\,1102+2054 is clearly much larger than one,
which makes this star unique among the many thousands of known white
dwarfs.

Because, according to the photometry, the effective temperature for
the two white dwarfs cannot be very different, the carbon abundance in
SDSS\,0922+2928 must be higher in order to produce the strong C$_2$
Swan bands. On the other hand, the hydrogen abundance must be lower,
because no Balmer lines are visible. We fixed the H/He ratio at
$10^{-5}$, the upper limit allowed by H$\beta$. Further reduction did
not substantially influence the models because the major electron
donors are C and O. Using $T_\mathrm{eff}=8270$\,K from the fit to the
photometry, and assuming $\log g=8.0$, we obtained a good fit to the
spectrum with the oxygen abundance decreased by $\sim0.2$\,dex and a
carbon abundance increased by $\sim0.6$\,dex with respect to
SDSS\,1102+2054 (Table\,3). Therefore, also in SDSS\,0922+2928,
$\mathrm{O/C}>1$. With the exception of the strong O\,I multiplets 778
and 845\,nm most of the higher excitation O\,I lines and the C\,I
lines show the same problem as in SDSS\,1102+2054: they are too weak
and too broad. This is again much improved at a hotter temperature,
e.g. 10000\,K, which is, however, in conflict with the photometry
(SOM, Sect.\,1).

The low H abundance suggests that SDSS\,0922+2928 and SDSS\,1102+2054,
similar as the other classes of hydrogen-deficient white dwarfs,
underwent a late shell-flash, leaving a He-dominated atmosphere. At
the low temperatures of these stars, the He convection zone extends
sufficiently deep to dredge core material up into the atmospheres
\cite{koesteretal82-2, pelletieretal86-1}. Given the large age of the
two stars ($>500$\,Myr), gravitational diffusion will unavoidably lead
to a larger concentration of carbon in the envelope. The only
plausible explanation for the observed O/C abundances is that these
two white dwarfs have overall very low carbon mass fractions, and
hence represent naked ONe cores. As such, they are  distinct
from 1H1504+65 and the ``hot'' DQ white dwarfs.

Most stellar models that produce ONe cores predict a layer of CO
surrounding the core that should be sufficiently thick to avoid upward
diffusion of large amounts of oxygen. However, a sequence of stellar
evolution calculations approaching the mass boundary of stars forming
ONe white dwarfs and those undergoing electron-capture SN\,II show
that the mass of the CO layer decreases with increasing initial
stellar mass \cite{garcia-berro+iben94-1, garcia-berroetal97-1,
  ibenetal97-1}. Thus SDSS\,0922+2928 and SDSS\,1102+2054 may have
descended from the most massive stars avoiding core-collapse, in which
case they would be expected to be very massive themselves. Our current
data is insufficient to provide any unambiguous measure of the masses
of SDSS\,0922+2928 and SDSS\,1102+2054, however their C/He abundance
is at the top of the range observed in the carbon-rich sequence of
"cool" DQ \cite{dufouretal05-1, koester+knist06-1} white dwarfs, some
of which have parallax measurements implying that they are of high
mass.

We explored spectral models for SDSS\,1102+2054 with the abundances
kept fixed as in Table\,3, but adopting surface gravities of $\log
g=8.5$ and 9.0, which correspond to masses of $0.9\,M_\odot$ and
$1.2M_\odot$ (SOM, Sect.\,2). Broadly similar fits to the absorption
lines can be achieved for higher surface gravities if the temperature
is increased by 1000 to 2000\,K as well. For $\log g=9.0$, the
strongest O\,I lines become somewhat too broad compared to the
observations, and we conclude that the currently available data is
consistent with masses of up to $\sim1M_\odot$.

Initial models of the evolution of intermediate mass stars predict
that ONe cores should also contain significant amounts of
magnesium. Updated nuclear reaction rates have lead to a substantial
downward revision of the Mg abundances \cite{gutierrezetal05-1}. For
SDSS\,1102+2054, which has the better quality spectrum of the two
white dwarfs presented here, we can place an upper limit on the
magnesium abundance of $\log[\mathrm{Mg/He}]<-6.1$ from the absence of
the Mg\,II 448\,nm line.

\bibliographystyle{Science}

%
%


\begin{scilastnote}
\item JG has been supported by an STFC studentship. BTG, TRM, and DS
  have been supported by an STFC rolling grant. We thank
  D. Townsley for useful discussions. Funding for the SDSS and SDSS-II
  has been provided by the Alfred P. Sloan Foundation, the
  Participating Institutions, the National Science Foundation, the
  U.S. Department of Energy, the National Aeronautics and Space
  Administration, the Japanese Monbukagakusho, the Max Planck Society,
  and the Higher Education Funding Council for England. The SDSS Web
  Site is http://www.sdss.org/. 
\end{scilastnote}


\clearpage

\centerline{\includegraphics[width=10cm]{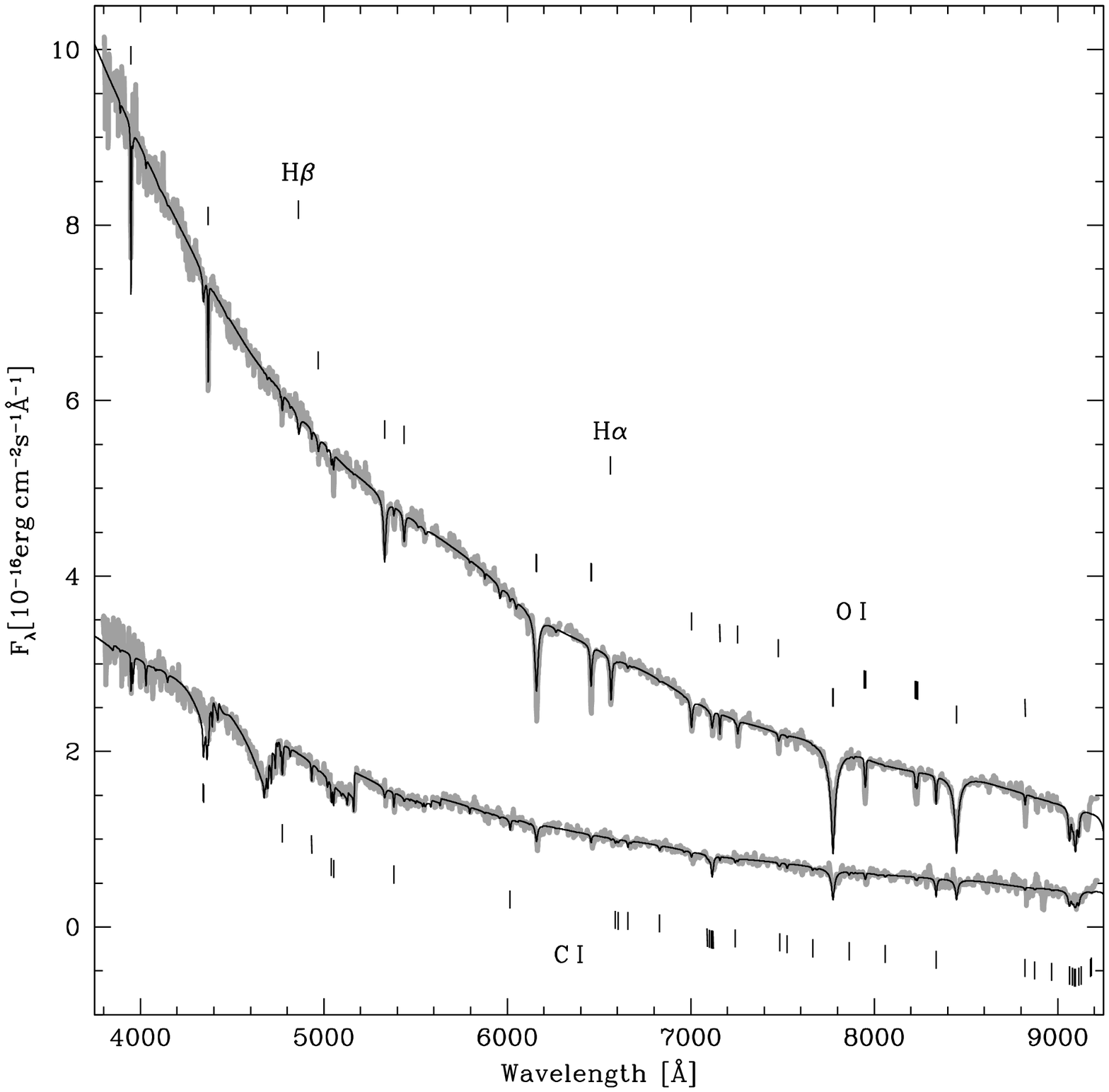}}

\medskip
\noindent 
{\bf Fig.\,1. Sloan Digital Sky Survey spectroscopy.}  The observed
spectra of SDSS\,0922+2928 (bottom) and SDSS\,1102+2054 (top)
are shown as thick gray lines. Superimposed as thin black lines
are models for $T_\mathrm{eff}=10000$\,K and $\log g=8$ (bottom)
and $T_\mathrm{eff}=10500$\,K and $\log g=8$ (top). The
wavelengths of the strongest OI and CI absorption lines, as well as of
H$\alpha$ and H$\beta$ are indicated by tick marks. 

\clearpage

\noindent {\bf Table 1. Equatorial coordinates, magnitudes and proper
  motions.}\\[1ex]
\begin{tabular}{lcc}
\hline
& SDSS\,0922+2928 & SDSS\,1102+2054 \\
\hline
RA(2000)  & 09 22 08.19    &  11 02 39.69   \\
Dec(2000) & +29 28 10.9    &  +20 54 39.4   \\
$u$       & $18.77\pm0.02$ &  $17.58\pm0.02$\\ 
$g$       & $18.52\pm0.02$ &  $17.24\pm0.01$\\
$r$       & $18.47\pm0.02$ &  $17.26\pm0.01$\\ 
$i$       & $18.58\pm0.02$ &  $17.34\pm0.01$\\
$z$       & $18.71\pm0.03$ &  $17.46\pm0.02$\\
p.m.      & 0.275"\,yr$^{-1}$ & 0.163"\,yr$^{-1}$\\
\hline                                                                  
\end{tabular}

\bigskip
\noindent {\bf Table 2. Atmospheric parameters for a surface gravity
  of $\log g=8$.}\\[1ex]
\begin{tabular}{lcccc}
\hline
System & 
  \multicolumn{2}{c}{$T_\mathrm{eff}$\,[K]} & 
  \multicolumn{2}{c}{distance\,[pc]} \\
 & pure He & with metals & pure He & with metals \\
\hline
SDSS\,0922+2928 & $8720\pm260$ & $8270\pm320$ & $141\pm6$ & $122\pm7$\\
SDSS\,1102+2054 & $8820\pm110$ & $8150\pm150$ & $81\pm2$ & $67\pm2$ \\
\hline                                                                  
\end{tabular}

\bigskip
\noindent {\bf Table 3. Photospheric abundances (by number).}\\[1ex]
\begin{tabular}{lcccc}
\hline
System ($T_\mathrm{eff}/\log g$)& $\log$(H/He) & $\log$(C/He) & $\log$(O/He) & $\log$(O/C) \\
\hline
SDSS\,1102+2054 (8150/8.0)  & -3.2 &  -3.2 & -1.8 & 1.4 \\
SDSS\,1102+2054 (10500/8.0) & -4.0 &  -3.6 & -1.8 & 1.8 \\
SDSS\,0922+2928 (8270/8.0)  & -5.0 &  -2.6 & -2.0 & 0.6 \\
\hline                                                                  
\end{tabular}

\newpage

\begin{center}
\Large
\textit{Supporting Online Material for}\\
Two white dwarfs with oxygen-rich atmospheres

\bigskip
\large{B.T. G\"ansicke$^1$, D. Koester$^2$,
J. Girven$^1$, T.R. Marsh$^1$, D. Steeghs$^1$}

\medskip
\normalsize
$^{1}$ Department of Physics, University of Warwick, Coventry CV4 7AL,
UK

\medskip
\normalsize
$^{2}$ Institut f\"ur Theoretische Physik und Astrophysik, University of Kiel,
24098 Kiel, Germany
\end{center}

\baselineskip16pt

\noindent
We describe here in more detail the white dwarf model atmospheres that
we have computed for the analysis of the SDSS data of SDSS\,0922+2928
and SDSS\,1102+2054, and outline the approach of our spectral
fits. Two grids of atmosphere models were calculated. The first grid
consisted of pure helium models, used to provide a first estimate of
the effective temperatures of both white dwarfs from fitting their
SDSS photometry. The second grid, including H, C, and O, was used to
refine the temperature estimate and to determine the atmospheric
abundances from fitting the observed absorption lines in the SDSS
spectra.

\section{He-model fits to the SDSS photometry}

In order to determine a first estimate of the effective temperatures
of the two white dwarfs, we fitted the He-models to the observed
$ugriz$ magnitudes, which were corrected to the AB scale by using the
corrections suggested by the SDSS project (-0.04, 0.00, 0.00, 0.01,
0.02)\footnote{http://www.sdss.org/dr7/algorithms/fluxcal.html\#sdss2ab}.
Because the mass of the two white dwarfs is currently unconstrained by
observations, we fixed the surface gravity to the canonical value of
$\log g=8.0$.  Good fits were achieved for both objects, implying
temperatures in the range 8000--9000\,K (Table\,2, Fig.\,S1). We note
that for both objects the SDSS data exhibits a discrepancy between the
$u$-band flux implied by the photometry, and the extrapolation of the
flux-calibrated spectroscopy. As part of our search for white dwarfs
with oxygen-rich white dwarfs, we have inspected the SDSS data of a
large number of objects, and find that such a discrepancy is rare, but
not truly exceptional, and based on the available data, it is not
possible to give preference to either the photometry or
spectroscopy. Taking the two spectral energy disributions at face value, a
fit to the photometry is bound to result in a lower temperature
compared to a fit to the spectroscopy. We will come back to this point
below.

\section{Fitting the SDSS spectroscopy with variable H,C, and O abundances}

Since He is practically neutral at the temperatures estimated above,
the three other elements (H, C, O) provide a significant electron
density, which correspondingly changes the opacity and thermal
structure of the atmospheres. The equation of state included 6
diatomic molecules (H$_2$, CH, OH, C$_2$, CO, O$_2$). In particular
the very stable molecule CO is important, because it reduces the
atomic carbon abundance considerably.  Also included in the second
grid was the continuous absorption of C and O using cross sections
obtained from the TOPBASE database of the Opacity Project
\cite{cuntoetal93-1, badnelletal05-1}.  The necessary O\,I and C\,I
line data were obtained from the VALD database \cite{piskunovetal95-1}
and from the Kurucz list (CDROM23). Many of the O\,I lines from Kurucz
list that were prominent in the model were not visible or much weaker
in the observed spectra, indicating that some of this line data is
wrong. We therefore gave preference to the VALD data, where the
problematic lines from the Kurucz list are either absent, or have $gf$
values several orders of magnitude smaller.  Recent calculations
\cite{celik+ates07-1} confirm the  data for atomic oxygen in the
databases we used.

There are  a number of O\,I lines where the core electrons are
not in the ground state ${}^4S{}^\mathrm{o}$ configuration, but in the
${}^2D{}^\mathrm{o}$ or ${}^2P{}^\mathrm{o}$ state. In these cases the core
electrons (without the excited electron which makes the transition)
are not in the configuration of the O\,II ground state and the
ionization therefore would occur to an excited level of the O\,II
ion. This has to be taken into account in our approximate calculation
of the van der Waals broadening constants, which depend on the
distance of the levels to the ionization limit. We use the
ionization limit of that series, that is 16.942 or 18.636\,eV instead
of the normal 13.618\,eV. From that limit and the term levels we
estimate the broadening constant using standard approximations
\cite{unsoeld55-1}. Similar approximations are also used for
the Stark broadening constants.

We re-fitted the SDSS $ugriz$ photometry with H, C, and O abundances
close to the final values, which resulted in slightly lower
temperatures for the two stars compared to the pure He-models
(Table\,2). In a final step, we varied the abundances of H, C, and O
while keeping the effective temperatures fixed (8270\,K for
SDSS\,0928+2928 and 8150\,K for SDSS\,1102+2054) and assuming $\log
g=8.0$. In this process, we focussed on the strong O\,I lines at 778
and 845\,nm, which originate from the lowest lying levels of all
optical lines and which only mildly depend on the prevailing effective
temperature. It turned out that no abundance pattern would provide a
fully satisfying fit to all observed lines, and we consider the
abundances listed in Table\,3 to represent the best compromise. 

A specific problem with this model is that the O\,I lines from the
higher levels are much too weak. One way to reduce the pressure in the
atmosphere models is to reduce the amount of He. We have calculated a
series of models between 7000 and 11000 K with the O/He ratio
increased up to values $>10$. In all cases the O\,I lines are much too
strong, and we rule out this possibility. The most likely solution to
this problem is then to assume a higher temperature, which increases
the strength of the high-excitation lines compared to the strong
multiplets. Assuming, as before, a canonical $\log g = 8$, a
reasonably good fit to the absorption lines was achieved with
$T_\mathrm{eff}=10500$\,K (Fig.\,1). However, the slope of this model
is substantially bluer than the SDSS $ugriz$ photometry (Fig.\,S1). A
fundamental uncertainty in modelling white dwarf spectra with such
unusual abundances is our limited knowledge on the sources of
ultraviolet opacity in their atmospheres. Any flux absorbed at
ultraviolet wavelengths will be re-distributed to longer wavelengths,
and hence result in a steepening of the slope in the optical
bandpass. An example of the marked differences in the amount of
ultraviolet opacity between pure He and C atmospheres has been
illustrated by \cite{somdufouretal08-1}. Our models predict strong
ultraviolet absorption edges of O\,I and C\,I (which we include with
the values obtained from TOPBASE), and a strong L$\alpha$ line
broadened mostly by neutral helium (which we describe with the theory
developed by our group). The currently available data does not provide
quantative constraints on the ultraviolet characteristics of
SDSS\,0922+2928 and SDSS\,1102+2054, and hence a systematic
uncertainty on their effective temperatures remains. We conclude that,
within the framework of the spectroscopic analysis, it is reasonable
to prefer the spectroscopic solution of the higher effective
temperatures, as it is impossible to satisfactorily model the observed
ratio of O\,I lines with high and intermediate excitation levels at
the low (photometric) temperature.

As discussed in the main manuscript, the large O/C abundance ratios
found for these two white dwarfs strongly suggests that they represent
bare oxygen-neon cores that may have descended from massive progenitor
stars. Under this assumption, their masses based on any of the current
initial mass-final mass relations \cite{catalanetal08-2,
  kaliraietal08-1, salarisetal09-1, casewelletal09-1, dobbieetal09-1,
  somwilliamsetal09-1} are expected to be $\sim1\,M_\odot$, corresponding
to a likely range for the surface gravity of $8.5\le\log g\le9.0$. In
order to explore the effect of a higher gravity on the spectral fit,
we have calculated a grid of models spanning the range $\log g=8.0$ to
9.0. Since the spectrum is largely dominated by the ionization balance
of He and the other elements, a change of $\log g$ can to some extent
be compensated by a change in temperature, and to first order the
abundances do not change significantly. In fact, we do find that
broadly similar fits can be achieved when increasing $T_\mathrm{eff}$
along with $\log g$, but keeping the abundances fixed at values given
in Table\,3.  Figure\,S2 illustrates the fits with our canonical model
($T_\mathrm{eff}=10\,500$\,K, $\log g=8.0$), and models for higher
surface gravities, ($T_\mathrm{eff}=11\,500$\,K, $\log g=8.5$) and
($T_\mathrm{eff}=12\,500$\,K, $\log g=9.0$). For $\log g=9.0$, the
strongest O\,I lines become somewhat too broad compared to the
observations. We conclude that the currently available data are
consistent with a broad range of white dwarf masses, including
relatively massive ($\sim1\,M_\odot$) stars, and that the main
conclusion of a large O/C abundance ratio is independent of the mass
assumed for the spectral fitting.


\begin{thebibliography}{10}
\bibitem{note1} We inspected a representative sub-set of all the DR7
  spectra in our sample and estimate that our analysis comprised
  $\simeq6000$ genuine white dwarfs. Applying the same color
  constraints used in our query on DR7 to the DR4 white dwarf
  catalogue \cite{eisensteinetal06-1} left $\simeq2400$ objects, which
  implies that using DR7 more than doubles the available candidate
  sample.
\end{thebibliography}
\bibliographystyle{Science}

\clearpage

\centerline{\includegraphics[width=12cm]{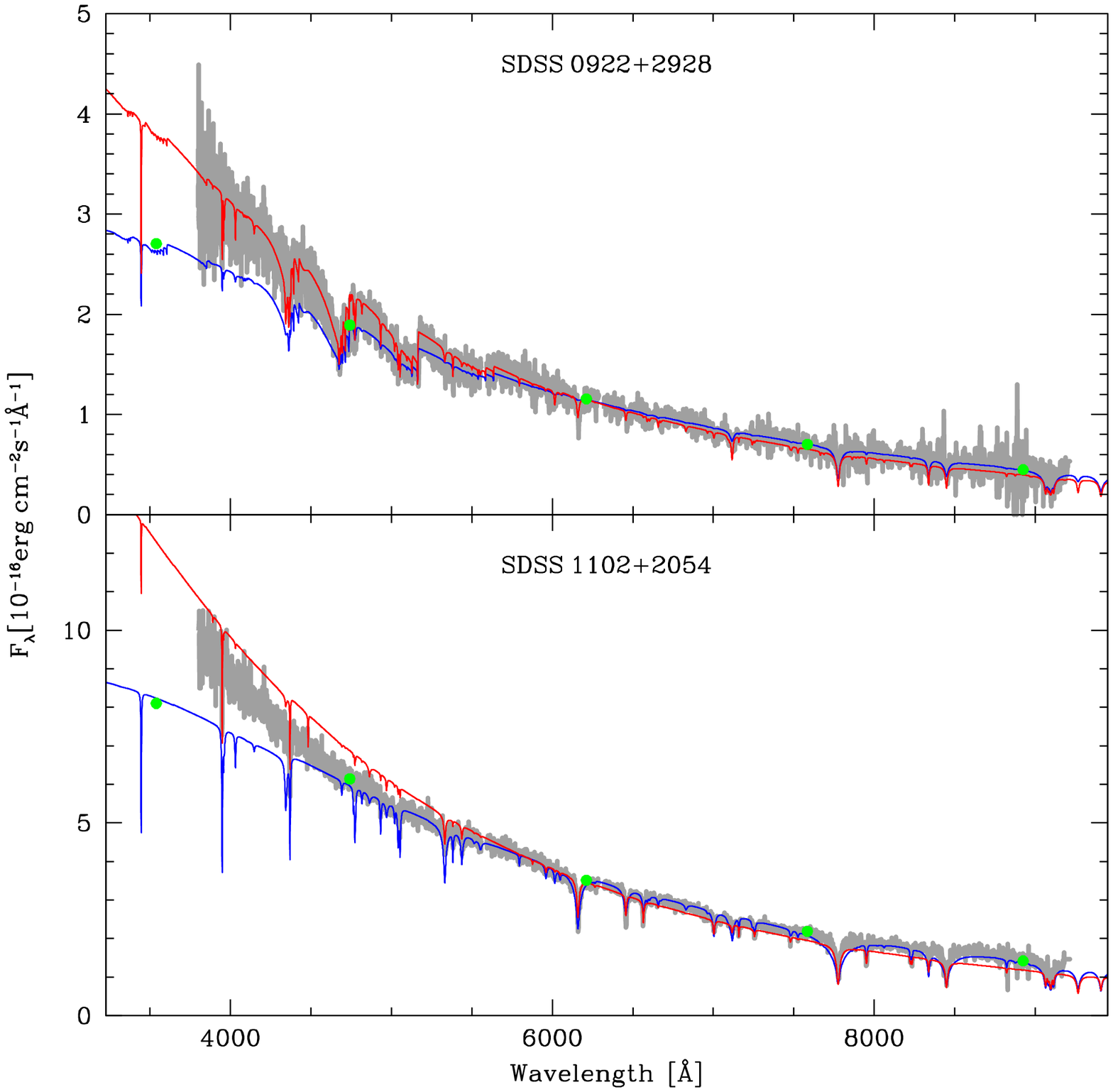}}
\bigskip
\noindent
\textbf{Supporting Figure\,1.} SDSS spectroscopy (gray lines) and
photometry (green dots) of SDSS\,1102 +2054 (bottom panel) and
SDSS\,0922+2928 (top panel). The uncertainties in the fluxes from the
SDSS photometry are comparable to the size of the dots. In both cases,
the $u$-band flux from the photometry lies below the extrapolation
of the spectroscopy.  The best-fit models to the photometry for an
assumed $\log g=8.0$ are shown as blue lines ($T_\mathrm{eff}=8270$\,K
for SDSS\,0922+2928 and $T_\mathrm{eff}=8150$\,K for SDSS\,1102+2054;
see Tables\,2 and 3). Shown as red lines are the best-fit models to
the absorption lines for an assumed $\log g=8.0$
($T_\mathrm{eff}=10\,000$\,K for SDSS\,0922+2928 and
$T_\mathrm{eff}=10\,500$\,K for SDSS\,1102+2054). As part of the
fitting the absorption lines, the models were normalised to the
observed SDSS spectra. Here, we show the best-fit models scaled to the
$r$-band magnitude without that normalisation to illustrate the
systematic uncertainty in the optical slope/effective temperature.

\centerline{\includegraphics[width=12cm]{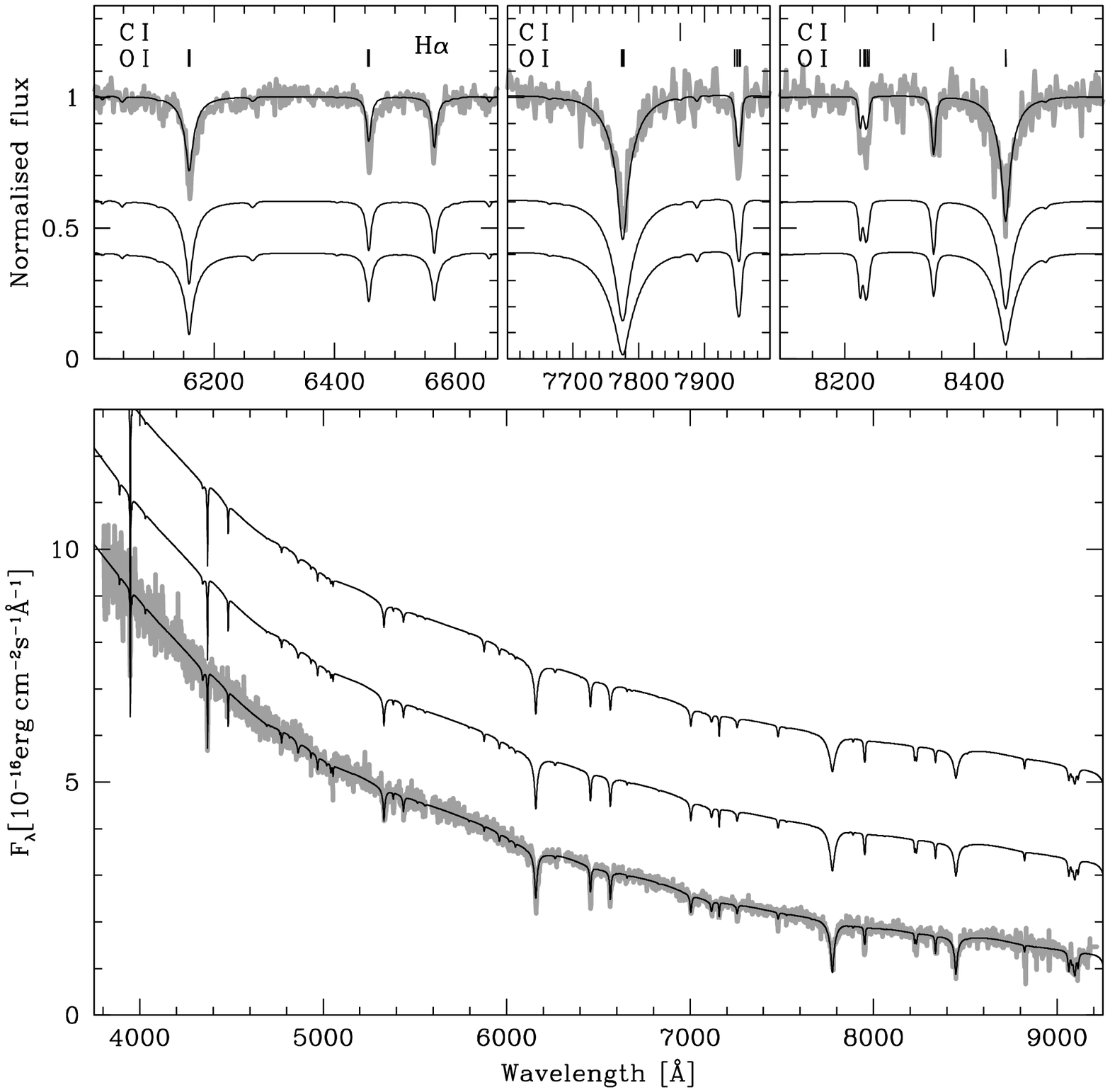}}
\bigskip
\noindent
\textbf{Supporting Figure\,2.} Model fits to the observed absorption
lines of SDSS\,1102+2054, illustrating the effect of varying the
surface gravity. The central panel shows the full wavelength range of the SDSS
spectroscopy, with a model for ($T_\mathrm{eff}=10\,500$\,K, $\log
g=8.0$) and the corresponding abundances from Table\,3 superimposed on
the SDSS data. Note that the continuum slope is normalised to the
observed data, and that only the absorption lines are fitted. Offset by
two and four flux units are spectra with ($T_\mathrm{eff}=11\,500$\,K,
$\log g=8.5$) and ($T_\mathrm{eff}=12\,500$\,K, $\log g=9.0$),
respectively, keeping the abundances fixed to those given in
Table\,3. The top panels show close-ups of the normalised SDSS
spectrum around several O\,I and C\,I lines, as well as of
H$\alpha$. Line identifications are indicated by tickmarks. A model
for ($T_\mathrm{eff}=10\,500$\,K, $\log g=8.0$) is superimposed on the
observed spectrum, models with ($T_\mathrm{eff}=11\,500$\,K, $\log
g=8.5$) and ($T_\mathrm{eff}=12\,500$\,K, $\log g=9.0$) are offset by
$-0.4$ and $-0.6$ flux units, respectively.

\end{document}